# Unravelling the Catalytic Activity of Dual-Metal Doped $N_6$-Graphene for Sulfur Reduction via Machine Learning-Accelerated First-Principles Calculations

Sahil Kumar[1], Adithya Maurya K R[1] and Mudit Dixit[1,2*]

[1]CSIR-Central Leather Research Institute (CSIR-CLRI), Chennai - 600020, India

[2]Academy of Scientific and Innovative Research (AcSIR), Ghaziabad 201002, India

Corresponding author email: *muditdixit@clri.res.in; dixitmuditk@gmail.com

## Abstract

Understanding and optimizing polysulfide adsorption and conversion processes are critical to mitigating shuttle effects and sluggish redox kinetics in lithium-sulfur batteries (LSBs). Here, we introduce a machine-learning-accelerated framework, Precise and Accurate Configuration Evaluation (PACE), that integrates Machine Learning Interatomic Potentials (MLIPs) with Density Functional Theory (DFT) to systematically explore adsorption configurations and energetics of a series of $N_6$-coordinated dual-atom catalysts (DACs). Our results demonstrate that, compared with single-atom catalysts, DACs exhibit improved LiPS adsorption and redox conversion through cooperative metal-sulfur interactions and electronic coupling between adjacent metal centers. Among all DACs, Fe-Ni and Fe-Pt show optimal catalytic performance, due to their optimal adsorption energies (-1.0 to -2.3 eV), low free-energy barriers (≤0.4 eV) for the $Li_2S_2$ to $Li_2S$ conversion, and facile $Li_2S$ decomposition barriers (≤1.0 eV). To accelerate catalyst screening, we further developed a machine learning (ML) regression model trained on DFT-calculated data to predict the Gibbs free energy (ΔG) of $Li_2S_n$ adsorption using physically interpretable descriptors. The Gradient Boosting Regression (GBR) model yields an $R^2$ of 0.85 and an MAE of 0.26 eV, enabling the rapid prediction of ΔG for unexplored DACs. Electronic-structure analyses reveal that the superior performance originates from the optimal d-band alignment and S-S bond polarization induced by the cooperative effect of dual metal centres. This dual ML-DFT framework demonstrates a generalizable, data-driven design strategy for the rational discovery of efficient catalysts for next-generation LSBs.

## Introduction

Optimizing lithium-ion batteries (LIBs) to achieve long driving ranges for pure electric vehicles (PEVs)remains a major challenge.[1] Owing to the high theoretical capacity (1675mAh $g^{-1}$)[2], high energy density (2600 $WhKg^{-1}$), and low cost of sulfur, lithium-sulfur batteries (LSBs) have emerged as promising candidates for next-generation high-capacity batteries.[3] Despite these advantages, the practical deployment of LSBs is severely constrained by several intrinsic challenges. The electrically insulating nature of elemental sulfur ($S_8$) and the poor utilization of active materials due to the formation of insoluble lithiated products, such as $Li_2S_2$ and $Li_2S$, hinder efficient electrochemical performance. Additionally, the dissolution and migration of higher-order lithium polysulfides ($Li_2S_n$, n > 4) during charge–discharge cycles, so-called "shuttle effect"[4] (Figure 1), lead to active material loss, self-discharge, and rapid capacity fading. To overcome these challenges, various strategies have been developed to improve sulfur utilization and polysulfide confinement. Among them, a promising strategy is to utilize single-atom catalysts (SACs) consisting of isolated transition-metal atoms coordinated with nitrogen-doped carbon substrates. SACs offer large surface areas, high electronic conductivity, and strong metal-sulfur binding through interactions between the Lewis basic nitrogen dopant and the Lewis acidic lithium atoms in $Li_2S_n$ species. SACs enhance the adsorption energy and reduce the dissolution of higher-order polysulfides by effectively trapping sulfur species. These unique attributes enable SACs to adsorb soluble polysulfides and improve the kinetics of the sulfur reduction reaction (SRR).[5] For example, Ni-based single-atom catalysts have been shown to promote the reduction of $S_8$ to $Li_2S_6$[6] while Fe, Co and Ni sites in SACs have been shown to facilitate sulfur species trapping and redox kinetics through coupling between the d-orbitals of metal centers and adsorbed intermediates[7][4c, 8] Despite these disadvantages, the sluggish redox kinetics and weaker binding with high-order polysulfide of LSB still remain a challenge. Additionally, due to the limited tunability of SACs, their utilization for addressing the problems of SIBs remains limited. To address these issues, dual-atom catalysts (DACs) have recently garnered considerable attention. In contrast to SAC, DACs have two neighboring transition-metal atoms coordinated by heteroatoms. The presence of two metal centers provides broader and tunable active sites. In addition, introduces synergistic effects that enhance both adsorption strength and catalytic activity. The cooperative interaction between the adjacent metal atoms enables improved intermediate stabilization,[9] thereby promoting faster sulfur reduction reaction (SRR) kinetics. Furthermore, DACs offer chemical flexibility, as the combination of different transition metals allows for fine-tuning of adsorption energetics and electronic structure, which are critical for achieving balanced polysulfide binding and catalytic efficiency.[5e, 10]

In this work, we investigate a series of DACs using DFT calculations to examine how the variation of transition metals (TMs) affects their catalytic activities. By systematically screening a series of DACs, we identify highly efficient DACs that enhance SSR kinetics through a synergistic effect. To achieve this, we design symmetric cathode materials where both metals are surrounded by the same number of pyridine, pyrrolic, and graphitized nitrogen atoms.[5e] Additionally, the distance between the parallel metals varies due to the change in ionic radii and electronegativity of both metals. Since the redox mechanism of SRR is complex to establish, we analyze the relationship between binding energy and Gibbs free energy with various electronic structure (band centers and density of states, DOS) and non-electronic structure (bond lengths of Li-S, bond angles of Li-S-Li, and number of sulfur atoms in polysulfides) descriptors that dictate the kinetics of SRR. Our results suggest that binding energy is not the only criterion for understanding the kinetics of SRR; there are many other hidden properties that require in-depth examination to comprehend the sluggish kinetics of SRR. Additionally, to efficiently identify favourable adsorption geometries of polysulfides on catalyst surfaces, we developed an automated screening framework, Precise and Accelerated Adsorption Configuration Evaluation (PACE), using the Atomic Simulation Environment (ASE). This method screens a large number of adsorbate configurations using Machine learning interatomic potentials (MLIP, MACE-mp-0). High-energy and duplicate conformers are removed, and the generated surface–adsorbate configurations are initially evaluated using single-point energy calculations with a machine-learned interatomic potential (MLIP). The lowest-energy configurations are further refined through structural optimization, and the top-ranked optimized geometries are treated with DFT calculations. This screening of adsorbate configurations allows us to identify the lowest-energy adsorbate configuration, which is otherwise difficult due to the highly heterogeneous nature of catalysts and polysulfides. Overall, this work identifies a highly active catalyst for LSBs, develops a machine learning model that predicts the relationships between electronic structure and activity, and establishes a method for fast and accurate prediction of polysulfide binding configurations on DACs.

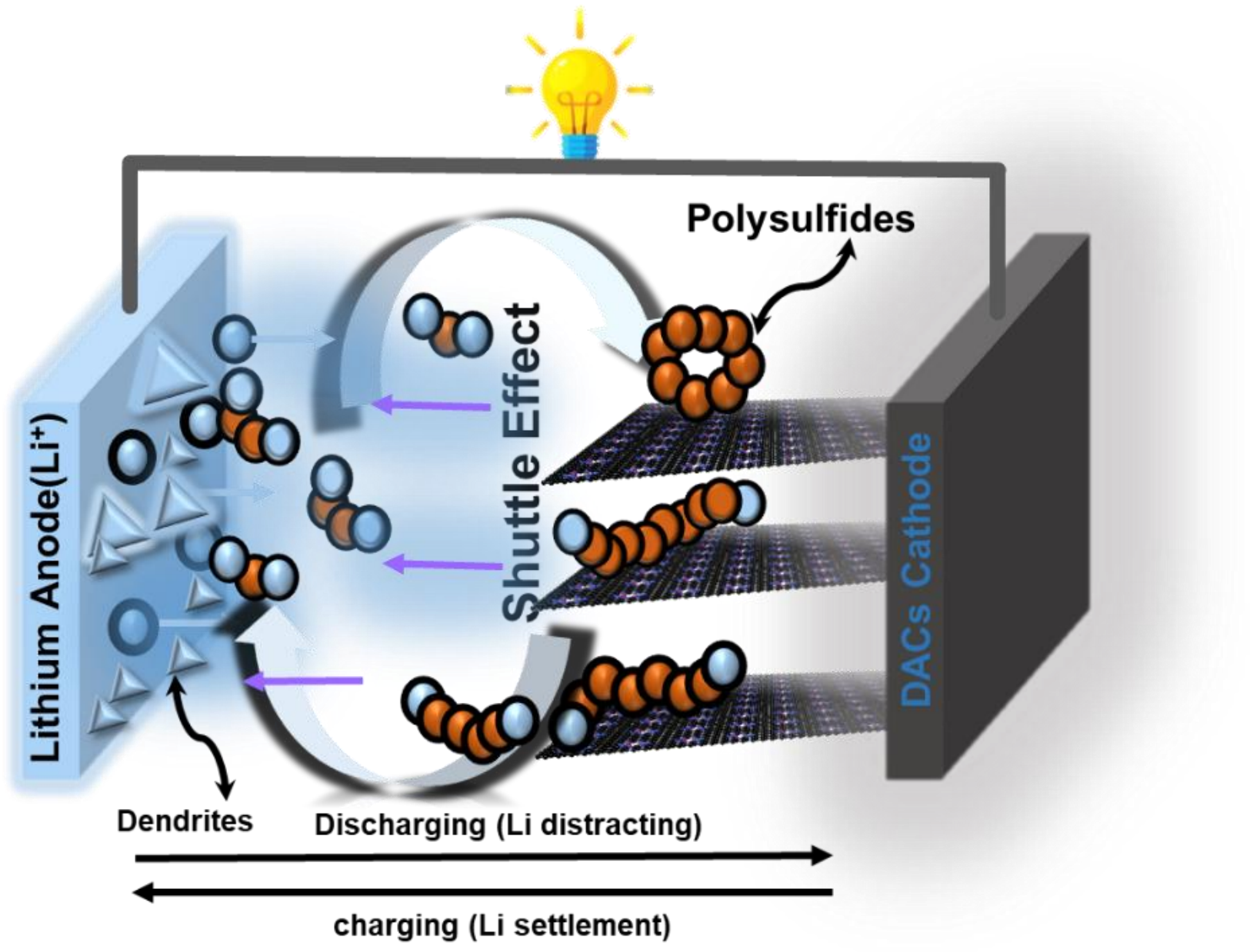

**Figure 1:** Schematic illustration of the reduction of higher order polysulfides to lower order ($Li_2S_n$ n = 1,2) and polysulfides shuttle effect in Li-S batteries.

## Results and Discussions

### Polysulfide Binding in DACs

During the discharging of LSBs, the elemental sulfur ($S_8$) gets converted to $Li_2S$ through a stepwise sulfur reduction reaction (SRR). The SSR begins with the conversion of $S_8$ into soluble higher-order lithium polysulfides ($Li_2S_n$, n = 4, 6, 8), followed by the formation of insoluble lower-order polysulfides ($Li_2S_n$, n = 1, 2). To understand the feasibility and efficiency of SRR on a given catalyst, the accurate assessment of the interactions involved in each intermediate is critical. A central challenge in modeling the SRR is to accurately identify the adsorption configuration of different polysulfide intermediates. Due to the high structural heterogeneity of catalysts and adsorbates, high-throughput screening of a large number of candidate structures with DFT could be computationally challenging for exhaustive screening, and the conventional approach to screen a few candidate structures with DFT may not guarantee the lowest energy structure of intermediates. In contrast, Machine Learning Interatomic Potentials (MLIPs) are compelling alternatives as these have shown a significant promise in predicting relaxed structures with reasonable accuracy while requiring only a fraction of the computational cost[11]. With the added advantage that MLIPs can parallelize computations over GPUs, it is possible to scale the number of screens to numbers beyond what traditional approaches can envision. To address this challenge, we have developed an in-house methodology for accelerated screening of adsorbate configurations and potential adsorption sites, named Precise and Accurate Adsorption Configuration Evaluation (PACE). This approach enables the rapid and systematic identification of low-energy adsorbate configurations on catalyst surfaces. PACE identifies stable ground state base-adsorbate configurations by first computing single-point MLIP calculations of different adsorbate configurations placed on predefined positions over the base. The initial placement coordinates of adsorbates are chosen from the intersection points of a uniformly spaced 3D grid constructed within the unit cell. The resolution of this grid, i.e., the number of subdivisions along each crystallographic axis, is a tunable parameter that allows control over the density of candidate adsorption sites. After different configurations of adsorbates on these sites are evaluated via the MLIP single-point calculations and ranked, a set of potential candidates is filtered, and MLIP structure optimization is performed on them. This is followed by first-principles DFT optimization on the MLIP predicted ground state energy system. This systematic evaluation for ground state configuration identification can provide accurate geometries of the intermediates that might be overlooked by other methods, especially for complex and heterogeneous adsorbate structures and catalysts with unusual adsorption sites. For instance, on a system of $N_6$-graphene doped with metals (Fe, Cu) and $Li_2S$ as adsorbate, 512 unique adsorbate configurations were used over different adsorption sites. This generates a total of 46,400 structures to filter from (with 10 subdivisions along a, b directions of crystallographic axis and at 2.5 Å above the slab). Using the MACE-MP-0 model[12], 40 MLIP structure optimizations were performed on the filtered systems and the predicted ground state system was used to identify the structure, as shown in Figure 2(a). A single NVIDIA V100 SXM2 GPU Card was used to run MLIP predictions. This approach has enabled us to identify potential adsorbed states across the studied systems, suggesting that it can be effectively employed for the systematic screening of ground-state configurations. As the MLIPs continue to improve in accuracy and generalizability, the effectiveness of this methodology correspondingly increases, enhancing its reliability for identifying low-energy adsorption configurations. Despite the screening, we further optimized these lowest energy structures with DFT using PBE and PBE+D3 functional for different systems (Figure S1-5).

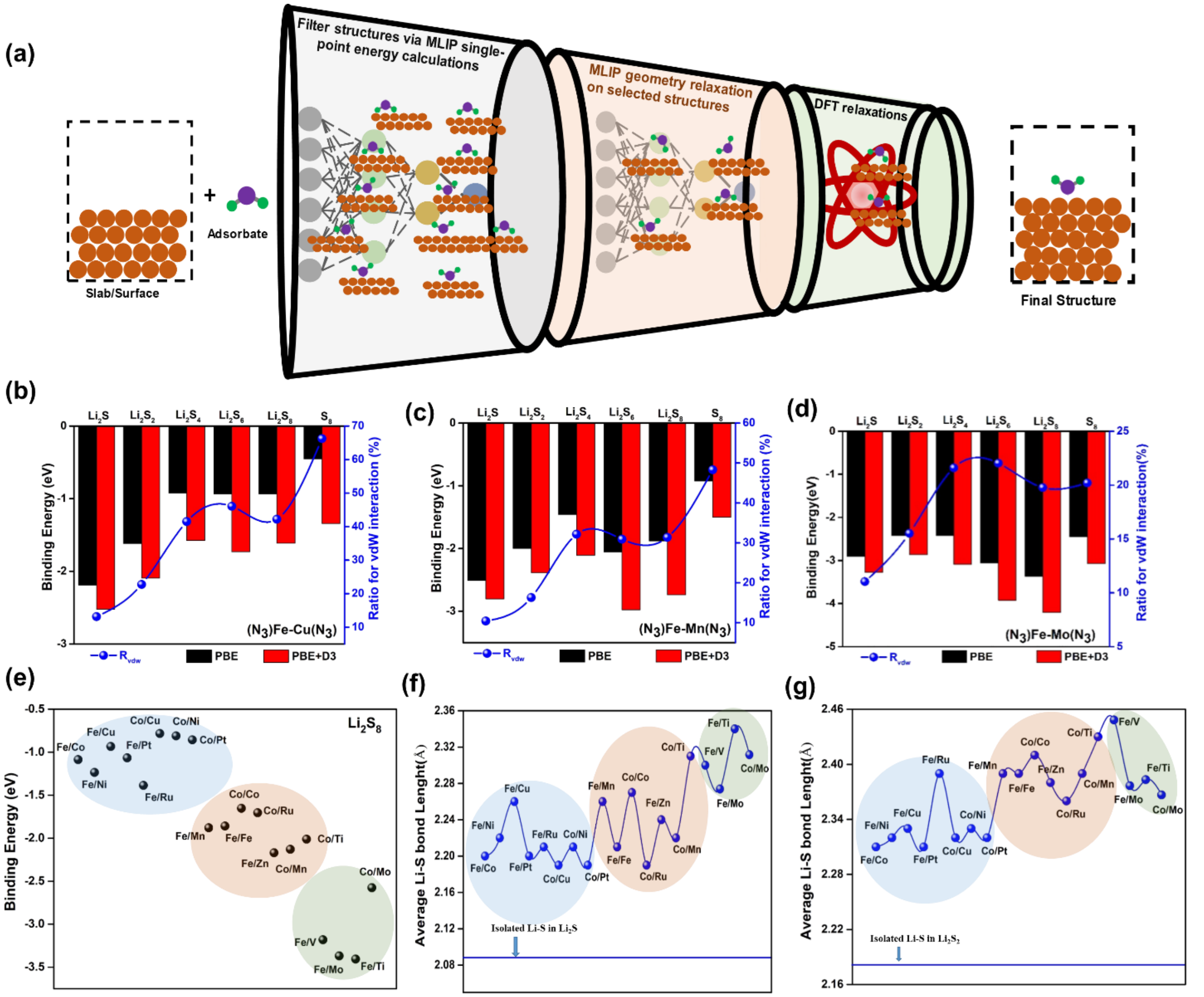

**Figure 2.** (a) Schematic diagram showcasing the working process of Precise and Accurate Adsorption Configuration Evaluation (PACE) for identifying the lowest energy structures of adsorbates. Binding energy of all polysulfides on N-doped graphene surface for (a) The Weak cathode hosts, (b) The Moderate cathode hosts, (c) The Strong cathode hosts (one example each). Black and red bars represent the PBE and PBE+D3 calculation data, and blue line represents the ratio of vdW, indicating the percentage difference between PBE and PBE+D3. (e) Binding energy of $Li_2S_8$ for all cathode hosts. Average Li-S bond length for all the cathode hosts in $Li_2S$(f) and $Li_2S_2$(g) compared to the isolated structure. Blue shade contains weak cathode hosts, Orange shade contains Moderate cathode hosts, and Greenish shade circle Strong Cathode hosts.

We first examined the binding energies of a representative high-order polysulfide, $Li_2S_8$, on single-atom catalysts (SACs) and dual-atom catalysts (DACs). We note that SACs ($Fe-N_4$ and $Co-N_4$) exhibit relatively weaker binding affinities, with binding energies of -0.61 eV ($Fe-N_4$) and -0.46 eV (Co-$N_4$) Figure S11 a,b. In contrast, DACs with nitrogen-coordinated bimetallic centers (($N_3$)Fe - M($N_3$))-0.8eV to -3.12 eV Figure 2(e), indicating that in DACs the sulfur-metal interaction and nitrogen-lithium interaction are relatively dominant compared to the SACs for effective anchoring of higher order polysulfides. A major issue in some Fe-M DACs is the extremely exothermic binding of polysulfides that can distort the DACs Figure S1-S5(q)-(s), resulting in strain in the system and thereby reducing the overall kinetics of DACs. To combat this issue, we carefully design a series of DACs by utilizing both the 3d and 4d metals with Fe and Co to tune their binding with higher-order polysulfides, ensuring the cathode material remains intact and effectively supports the SRR mechanism for long-term stability.

To further deconvolute the contribution of physical and chemical interactions in polysulfide adsorption, we compared the van der Waals interaction ratio[13] ($R_{vdw}$, Figure 2(b-d)), which reveals an almost consistent increase in van der Waals contributions from lower- to higher-order polysulfides. In lower-order polysulfides, the chemical interaction predominates due to the higher concentration of the Lithium atoms. Interestingly, on several DACs, $Li_2S_6$ exhibited stronger van der Waals interactions than $Li_2S_8$ Figure 3, which can be attributed to the higher structural flexibility and compactness of $Li_2S_6$ Figure S6-S8. Compared to DACs, SACs generally display weaker chemisorption and relatively higher $R_{vdw}$ values Figure S11 (a,b), reflecting the limited spatial extent of their active sites. This reduced interaction strength in SACs leads to relatively poor anchoring of higher-order polysulfides (HOPs), enhancing polysulfide shuttling and compromising redox performance. The analysis of the $Li_2S_8$-DACs systems revealed that there are two distinct modes of adsorption of HOPs on DACs. In the first mode, a sulfur atom bridges both metal centers (a sulfur atom of polysulfide interacts with both the metals at the same time, M-S distances, Table S2) in the DAC, as observed in systems such as ($N_3$)Fe–Co($N_3$) Figure 3(h) and $(N_3)$Fe–Ni$(N_3)$ Figure 3(i). In the second mode, the sulfur atoms of polysulfide interact primarily with a single metal atom (a sulfur atom of polysulfide interacts with a single metal, M-S distances, Table S2), evident in systems like $(N_3)$Fe–Cu$(N_3)$ Figure S5(c) and $(N_3)$Co–Pt$(N_3)$ Figure S5(h). Despite the single M-S coordination in some of the DACs, the binding energies of $Li_2S_8$ remain more exothermic than the SACs, due to the synergetic effect of both metals Figure S10. As expected, configurations involving dual-metal coordination exhibit stronger binding due to the cooperative effect on both metals. These observation highlights that in DACs, both metals participate in the binding of polysulfide, and there is a synergetic effect of both metals in binding. From a comparative analysis of vdW contribution and M-S bond lengths, it is evident that in DACs there is a larger extent of chemisorption than in SACs.

Based on the binding affinity of DACs with $Li_2S_8$, we categorized these into three sets of cathode hosts: (i) weak cathode host ($E_b$ = -0.7eV to -1.5eV), (ii) moderate cathode host ($E_b$ = -1.5eV to -2.5eV), and (iii) strong cathode host ($E_b$ more exothermic than -2.5 eV).

For instance, in the $(N_3)$Fe-Ru$(N_3)$, the Ru-S bond length (2.27Å, Table S2) is shorter than Fe-S (2.36Å), suggesting that Ru plays a dominant role in stabilizing the adsorbate. A similar trend is observed in $(N_3)$Fe-Co$(N_3)$, where the Co-S bond (2.25Å) is shorter than the Fe-S bond (2.44Å, Table S2) and in $(N_3)$Co-Ni$(N_3)$, where the Co-S bond (2.17 Å) is shorter than the Ni-S bond (2.30 Å), suggesting that in these systems Co plays a dominant role in binding.

On weak cathode hosts such as $(N_3)$Fe-Pt$(N_3)$ and $(N_3)$Fe-Ru$(N_3)$, the lower-order polysulfides ($Li_2S$ and $Li_2S_2$) exhibit higher binding affinities than higher-order polysulfides. Specifically, the binding energies for $Li_2S$ are -2.35 eV (Fe-Pt) and -2.35 eV (Fe-Ru), while for $Li_2S_2$ they are -1.72 eV and -1.83 eV, respectively. Among weak cathode hosts, Fe-based 3d catalysts show stronger adsorption of $Li_2S$ compared to their Co-based counterparts. For example, $(N_3)$Fe-Co$(N_3)$ Figure 3(b), $(N_3)Fe-Ni(N_3)$ Figure 3(c), and $(N_3)Fe-Cu(N_3)$ Figure 3(d) exhibit binding energies of -2.26 eV, -2.29 eV, and -2.19 eV, respectively, in contrast to $(N_3)Co-Cu(N_3)$ (-1.82 eV; Figure 3(e) and $(N_3)Co-Ni(N_3)$ (-1.98 eV; Figure S6(c)). This suggests that Fe-based catalysts bind sulfur more strongly because Fe can accept electrons more readily than Co due to the presence of vacant orbitals shown by difference charge density Figure 6(e). Furthermore, the stronger Fe-S interaction is favored by the soft-soft acid-base interaction principle. A similar binding affinity trend is observed for $Li_2S_2$. Moreover, we find that upon adsorption of $Li_2S$ and $Li_2S_2$ on DACs, the average Li-S bond length increases Figure 2(f-g), which could help in lowering the barrier for the formation of elemental sulfur. Now we turn our attention to moderate hosts. As expected, the binding affinity of $Li_2S_8$ on moderated cathode hosts is more exothermic than on weak hosts. For instance, on $(N_3)Fe-Mn(N_3)$, $(N_3)Fe-Fe(N_3)$ and $(N_3)Co-Co(N_3)$, the binding energies of $Li_2S_8$ are -1.87eV, -1.85eV and -1.65eV respectively Figure 2(e). Similarly, $Li_2S$ exhibits significantly stronger adsorption on moderate hosts than on weak hosts, suggesting that these dual-atom catalysts preferentially bind sulfur atoms, thereby facilitating $Li^+$ migration more effectively. Notably, the average Li-S bond length in $Li_2S$ adsorbed on DACs is shorter than that in $Li_2S_2$ (Figure 2(f-g)). This bond shortening indicates that the catalysts may facilitate Li-S bond dissociation, thereby promoting more efficient polysulfide conversion.

Further, for $(N_3)Co-Ru(N_3)$ Figures S1(m), S2(m), and $(N_3)Co-Ti(N_3)$ Figures S1(o), S2(o), the sulfur atom of $Li_2S$ and $Li_2S_2$ is singly bonded with one of the metals of given DAC, indicating a minimal synergistic effect. Although these systems still exhibit strong adsorption, the electron redistribution from both metals to sulfur Figure S16 is less pronounced for $(N_3)Co-Ti(N_3)$ compared to $(N_3)Fe-Mn(N_3)$ Figure S7(a), $(N_3)Fe-Fe(N_3)$ Figure 3(a), $(N_3)Co-Co(N_3)$ Figure

S7(c), and ($N_3$)Fe-Zn($N_3$) Figure S7(b), which exhibit stronger binding through bridge-sulfur bonding with both metals.

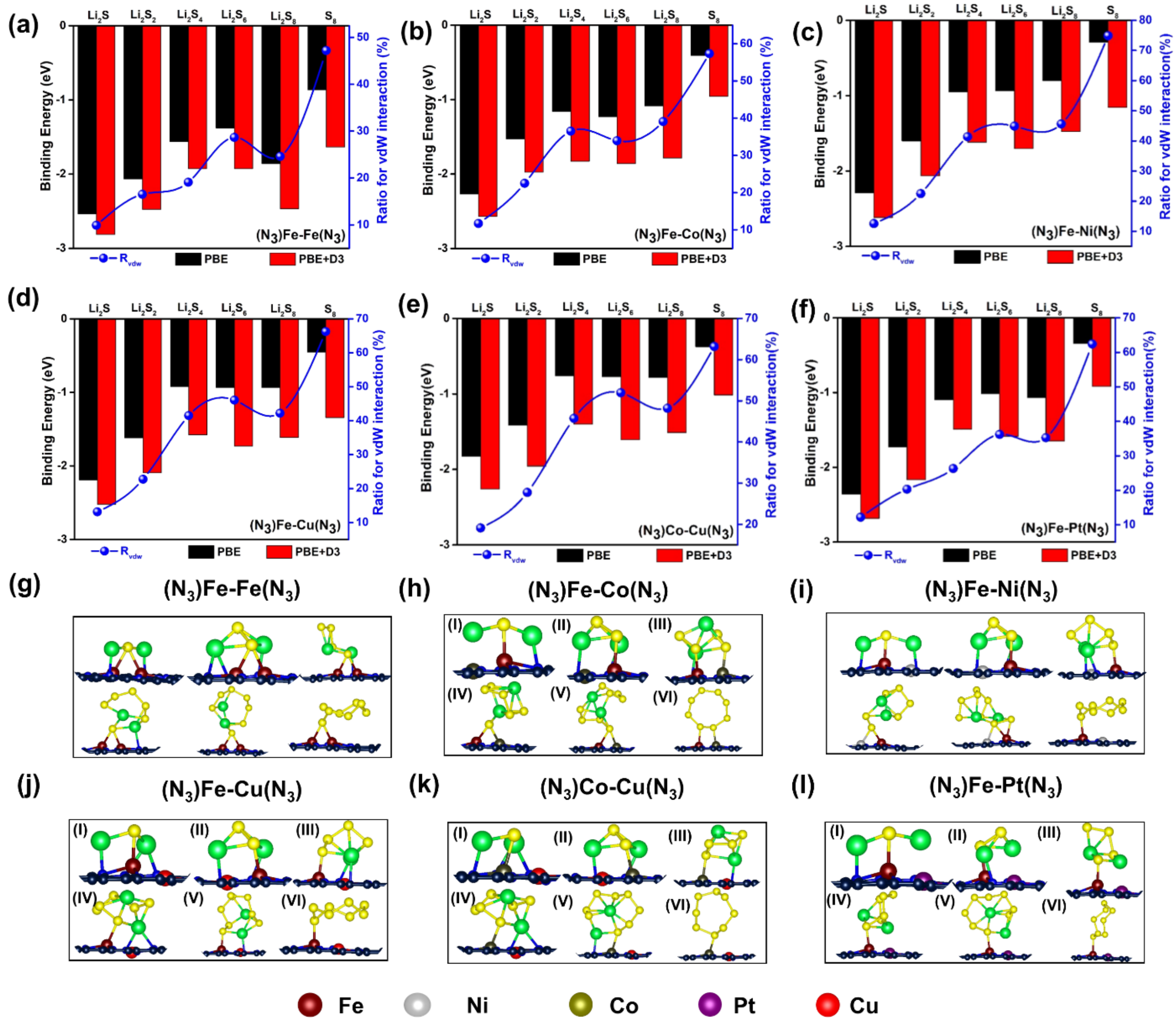

**Figure 3:** The Binding energies of polysulfides on dual metal N-doped graphene surface of (a) ($N_3$) Fe-Fe($N_3$), (b) ($N_3$)Fe-Co($N_3$), (c) ($N_3$)Fe-Ni($N_3$), (d) ($N_3$)Fe-Cu($N_3$), (e) ($N_3$)Co-Cu($N_3$), (f) ($N_3$)Fe-Pt($N_3$). Black and red bar represents the PBE and PBE+D3 calculation data, and blue line represents the ratio of vdW, indicating the percentage difference between PBE and PBE+D3. Structural side view after the adsorption of polysulfides on dual metal N-doped graphene surface, (g) ($N_3$) Fe-Fe($N_3$), (h) ($N_3$)Fe-Co($N_3$), (i) ($N_3$)Fe-Ni($N_3$), (j) ($N_3$)Fe-Cu($N_3$), (k) ($N_3$)Co-Cu($N_3$), (l) ($N_3$)Fe-Pt($N_3$). The inset images in panels (g) to (I) illustrate the optimised structures of $Li_2S$, $Li_2S_2$, $Li_2S_4$, $Li_2S_6$, $Li_2S_8$ and $S_8$, respectively.

In the case of $Li_2S_2$, the binding strength decreases significantly for $(N_3)$Fe-Zn$(N_3)$ because the metal and sulfur are singly bonded due to the filled d-orbital of Zn Figure S2(k). In contrast, for $(N_3)$Co-Co$(N_3)$, the difference in binding strength between $Li_2S$ and $Li_2S_2$ is nearly same due to the presence of bridges Figure S1-S2(l), sulfur bonds to both metals.

For strong cathode hosts, only $(N_3)$Fe-V$(N_3)$ Figures S1(p), S2(p) features a bridged sulfur bound to both the metal in the lower-order polysulfide. This configuration leads to strong adsorption energies, with average Li-S distances of 2.29 Å for $Li_2S$ Figure 2(f) Such bonding suggests that $(N_3)$Fe-V$(N_3)$ is an excellent candidate for promoting fast kinetics in the formation of elemental sulfur.

**Catalytic Activity of SRR**

As we discussed earlier, the redox mechanism of SRR is not solely governed by the binding energy of the polysulfides. Therefore, to gain deeper insight into the redox kinetics, we analyze the Gibbs free energy change from each elementary step involved in the formation of $Li_2S$ from $S_8$. The reaction of sulfur reduction is considered as follows[13-14]:

$$*\mathrm{S}_8 + 2\mathrm{Li}^+ + 2\mathrm{e}^- \longrightarrow *\mathrm{Li}_2S_8 \quad (1)$$

$$3*\mathrm{Li}_2S_8 + 2\mathrm{Li}^+ + 2\mathrm{e}^- \longrightarrow 4*\mathrm{Li}_2S_6 \quad (2)$$

$$2*\mathrm{Li}_2S_6 + 2\mathrm{Li}^+ + \mathrm{e}^- \longrightarrow 3*\mathrm{Li}_2S_4 \quad (3)$$

$$*\mathrm{Li}_2S_4 + 2\mathrm{Li}^+ + 2\mathrm{e}^- \longrightarrow 2*\mathrm{Li}_2S_2 \quad (4)$$

$$*\mathrm{Li}_2S_2 + 2\mathrm{Li}^+ + 2\mathrm{e}^- \longrightarrow 2*\mathrm{Li}_2S \quad (5)$$

Here * represents the DAC,

From the above equation, we calculate the change of Gibbs free energies by the equation written below:

$$\Delta G_1 = G\ (^*Li_2S_8) - G\ (*S_8) - 2G\ (Li) \tag{6}$$

$$\Delta G_2 = 4G\ (^*Li_2S_6) - 3G\ (*Li_2S_8) - 2G\ (Li) \tag{7}$$

$$\Delta G_3 = 3G\ (^*Li_2S_4) - 2G\ (*Li_2S_6) - 2G\ (Li) \tag{8}$$

$$\Delta G_4 = 2G\ (^*Li_2S_2) - G\ (*Li_2S_4) - 2G\ (Li) \tag{9}$$

$$\Delta G_5 = 2G\ (^*Li_2S) - G\ (*Li_2S_2) - 2G\ (Li) \tag{10}$$

For weak cathode hosts, the free energy profiles indicate that the initial reduction of $S_8$ to $Li_2S_8$ is highly exothermic across all systems. Among the 3d-based weak host DACs, due to the highest binding affinity of $Li_2S_8$ ($E_b$ = -1.23 eV), the free energy for the conversion of $S_8$ to $Li_2S_8$ is the most exothermic on $(N_3)$Fe-Ni$(N_3)$ with $\Delta G$ = -3.26 eV. Additionally, on $(N_3)$Fe-Co$(N_3)$, $(N_3)$Fe-Cu$(N_3)$ and $(N_3)$Co-Cu$(N_3)$ Figure 4 the change of Gibbs free for this step are -2.99 eV ($E_b$ = -1.08 eV), -2.77 eV($E_b$ = -0.94 eV), and -2.72 eV ($E_b$ = -0.78eV), respectively due to relatively less exothermic binding energies. Similarly, with 4d metal doped weak hosts $(N_3)$Fe-Ru$(N_3)$, $(N_3)$Co-Pt$(N_3)$ Figure S9(a) and $(N_3)$Fe-Pt$(N_3)$ Figure 4, and the Free energy of conversion from $S_8$ to $Li_2S_8$ is exothermic as $\Delta G$ are -2.43 eV, -2.80 eV and -3.01 eV, with respective binding energies of -1.38 eV, -0.85 eV and -1.06 eV, respectively. We noted that the conversion of $Li_2S_8$ to $Li_2S_6$ is exothermic on a few weak hosts, including $(N_3)$Fe-Co$(N_3)$, $(N_3)$Fe-Cu$(N_3)$, $(N_3)$Co-Ni$(N_3)$, and $(N_3)$Co-Pt$(N_3)$ as for these DACs, the binding affinity increases from $Li_2S_8$ to $Li_2S_6$. However, on $(N_3)$Fe-Ni$(N_3)$, $(N_3)$Co-Cu$(N_3)$, $(N_3)$Fe-Ru$(N_3)$, and $(N_3)$Fe-Pt$(N_3)$, binding affinity decreases from $Li_2S_8$ to $Li_2S_6$, resulting in the endothermic or nearly thermoneutral conversion (for $(N_3)$Co-Cu$(N_3)$ and $(N_3)$Fe-Pt$(N_3)$). Expectedly, for the conversion of $Li_2S_6$ to $Li_2S_4$, the change in Gibbs free energy is endothermic as the binding affinity decreases accordingly, except for $(N_3)$ Fe-Ni$(N_3)$ (thermoneutral) and $(N_3)$Fe-Ru$(N_3)$ (endothermic). The subsequent conversion of $Li_2S_8$ to $Li_2S_6$ is exothermic for several weak hosts, including $(N_3)$Fe-Co$(N_3)$, $(N_3)$Fe-Cu$(N_3)$, $(N_3)$Co-Ni$(N_3)$, and $(N_3)$Co-Pt$(N_3)$, owing to an increase in binding affinity. In contrast, $(N_3)$Fe-Ni$(N_3)$, $(N_3)$Co-Cu$(N_3)$, $(N_3)$Fe-Ru$(N_3)$, and $(N_3)$Fe-Pt$(N_3)$ show less exothermic binding energy change on moving from $Li_2S_8$ to $Li_2S_6$, resulting in endothermic or nearly thermoneutral free energies. The conversion of $Li_2S_6$ to $Li_2S_4$ is generally endothermic for most hosts, except for $(N_3)$Fe-Ni$(N_3)$ (thermoneutral) and $(N_3)$Fe–Ru$(N_3)$ (slightly endothermic) due to increased binding energy in $Li_2S_4$. The formation of solid polysulfides ($Li_2S_2$ and $Li_2S$) from $Li_2S_4$ is widely regarded as the rate-determining step (RDS) in the SRR pathway, consistent with our findings that this step is the most endothermic on $(N_3)$Fe-Co$(N_3)$ ($\Delta G$ = 0.64 eV), $(N_3)$Fe-Ni$(N_3)$ ($\Delta G$ = 0.42 eV), $(N_3)$Co–Ni$(N_3)$ ($\Delta G$ = 0.60 eV), and $(N_3)$Fe-Pt$(N_3)$ ($\Delta G$ = 0.50 eV). By contrast, $(N_3)$Fe-Cu$(N_3)$, $(N_3)$Co-Cu$(N_3)$, $(N_3)$Fe-Ru$(N_3)$, and $(N_3)$Co-Pt$(N_3)$ have the RDS in the final $Li_2S_2$ to $Li_2S$ conversion, with free energy changes of 0.46 eV, 0.60 eV, 0.59 eV, and 0.61 eV, respectively. Overall SRR free energies are highly exothermic for $(N_3)$Fe-Pt$(N_3)$ ($\Delta G$ = -1.98 eV), $(N_3)$Fe-Ni$(N_3)$ ($\Delta G$ = -1.95 eV), and $(N_3)$Fe–Co$(N_3)$ ($\Delta G$ = -1.82 eV), highlighting their potential as effective SRR catalysts. Notably, these hosts also display the least endothermic $Li_2S_2$ to $Li_2S$ conversion, with $\Delta G$ values of 0.39 eV, 0.36 eV, and 0.31 eV, respectively, compared with 0.46 eV for $(N_3)$Fe-Cu$(N_3)$, indicating faster kinetics for complete sulfur reduction.

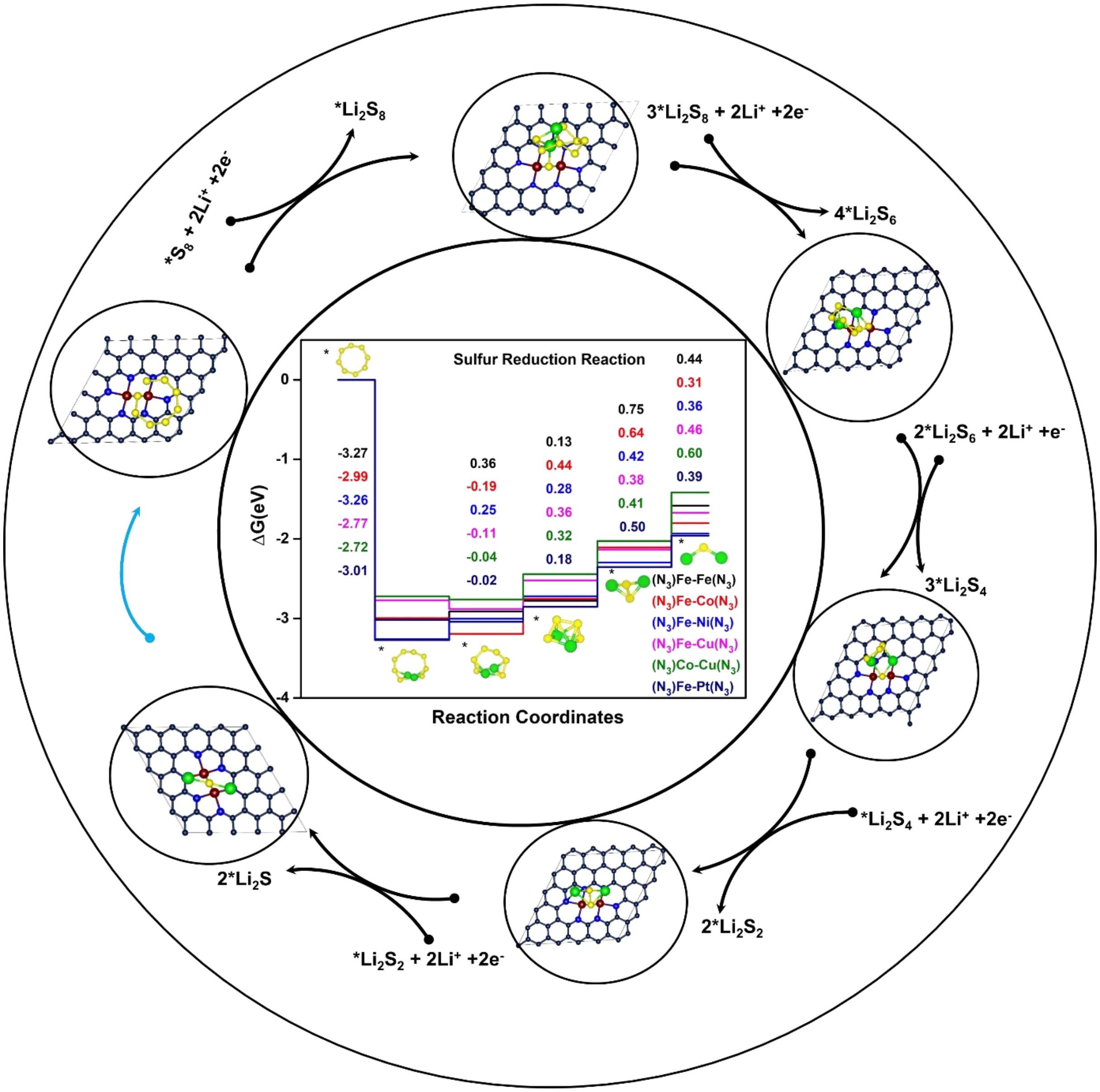

**Figure 4:** Gibbs Free energy landscape for all polysulfides of the sulfur reduction reaction on the dual metal N-doped graphene surface. The insets show the corresponding optimized structural configuration (Top view) of the moderate cathode host $(N_3)Fe\text{-}Fe(N_3)$ to represent the active site of the DACs.

Among the moderate cathode hosts, $(N_3)Fe\text{-}Zn(N_3)$, $(N_3)Co\text{-}Mn(N_3)$, and $(N_3)Co\text{-}Ti(N_3)$ exhibit the most exothermic Gibbs free energy changes for the conversion of $S_8$ to $Li_2S_8$, with values of -3.55 eV ($E_b$ = -2.17 eV), -3.45 eV ($E_b$ = -2.13 eV), and -3.49 eV ($E_b$ = -2.02 eV) Figure S9(b), respectively, arising from their highly exothermic $Li_2S_8$ binding energies. In contrast, $(N_3)$Co-Co($N_3$) ($E_b$ = -1.65 eV) Figure S9(b) and $(N_3)$Fe–Fe($N_3$) Figure 4 have relatively less exothermic $Li_2S_8$ binding energies, leading to smaller free energy changes, likely due to competition between the two metal centers in the electron transfer process, as indicated by similar M-S bond lengths for both metals in these DACs. Nevertheless, DACs such as $(N_3)$Fe-Mn($N_3$) Figure 4 and $(N_3)$Co–Ru($N_3$) Figure S9(b) still exhibit notably exothermic $S_8$ to $Li_2S_8$ free energy changes, possibly due to a strong synergistic effect and electron redistribution between the metals. The binding energy of $Li_2S_4$ on $(N_3)$Fe-Mn($N_3$) is relatively low ($E_b$ = -1.45 eV), making the $Li_2S_6$ to $Li_2S_4$ conversion step the RDS for this host. For systems such as $(N_3)$Co-Co($N_3$), $(N_3)$Co-Ru($N_3$), and $(N_3)$Co-Mn($N_3$) Figure S9(b), the RDS is the conversion of $Li_2S_2$ to $Li_2S$ conversion due to the strong adsorption energies of $Li_2S_2$ (-1.98 eV, -2.29 eV, and -2.04 eV, respectively). In the case of $(N_3)$Fe-Zn($N_3$), the final $Li_2S$ formation step is exothermic, which can be attributed to the dominant Fe-S interaction, as Fe forms stronger bonds with sulfur compared to Zn. Overall, among moderate hosts, $(N_3)$Fe-Fe($N_3$), $(N_3)$Co-Co($N_3$), and $(N_3)$Fe-Zn($N_3$) show higher overall Gibbs free energies compared to the weak cathode hosts. In the strong host category systems such as $(N_3)$Fe-V($N_3$), $(N_3)$Fe-Mo(N3), $(N_3)$Fe-Ti($N_3$), and $(N_3)$Co-Mo($N_3$) also exhibit large exothermic $S_8$ to $Li_2S_8$ free energy changes of -2.87 eV, -3.21 eV, -3.0 eV, and -2.42 eV, respectively Figure S9(c). However, due to the consistently high binding strength for all polysulfide intermediates in these materials, the overall Gibbs free energies are less favorable, with values of -0.84 eV, -0.44 eV, -0.66 eV, and -0.89 eV, respectively. The RDS is the $Li_2S_4$ to $Li_2S_2$ conversion phase transition step that limits their suitability as SRR catalysts. Although the conversion of $S_8$ to $Li_2S_8$ on strong cathode hosts ($(N_3)$Fe-V($N_3$), $(N_3)$Fe-Mo($N_3$), $(N_3)$Fe-Ti($N_3$), and $(N_3)$Co-Mo($N_3$)) is highly exothermic with $\Delta G$ values of -2.87 eV, -3.21 eV, -3.00 eV, and -2.42eV), extremely strong binding of polysulfides makes the RDS step ($Li_2S_4$ to $Li_2S_2$) highly endothermic and resulting in low overall Gibbs free energies thereby making then not potentially good candidates for SSR at ambient conditions.

**Electronic Structure Interpretation:**

After analyzing the adsorption energies and thermodynamics of all the polysulfides, we further examined the electronic structures of the catalysts and their corresponding $Li_2S_2$-adsorbed systems by investigating the projected density of states (PDOS). Since S-S bond dissociation is a key step in the sulfur reduction reaction (SRR), we focused on $Li_2S_2$ adsorption on DACs. The occupation and charge transfer between the d-orbitals of transition metals and the p-orbitals of sulfur play a crucial role in adsorption. To capture these effects, we compared the PDOS of pristine DACs with those of $Li_2S_2$-adsorbed systems Figure 5(a-b). For $(N_3)$Fe-Co$(N_3)$, $Li_2S_2$ adsorption increases the population of Co-d states below the Fermi level, indicating electron gain on Co. Conversely, the contribution of Fe-d states near the Fermi level decreases slightly, suggesting minor electron loss from Fe Figure 5(a-b). In the case of $(N_3)$Fe-Cu$(N_3)$, the population of Fe-d states significantly increases below the Fermi level, indicating electron gain upon $Li_2S_2$ adsorption, whereas the Cu-d states below the Fermi level are reduced, signifying electron loss from Cu. These observations highlight the synergistic electron redistribution that facilitates adsorbate binding during SRR. Interestingly, for $(N_3)$Fe–Ni$(N_3)$, we observe a noticeable increase in the d-state population of Ni near the Fermi level Figure 5(a-b), likely arising from electron gain and associated structural distortions upon $Li_2S_2$ binding. On pinpointing the electronic structure of $Li_2S_2$, we notice that upon adsorption, the electronic structure of both the sulfur atoms no longer remains the same (as seen in the molecular $Li_2S_2$ Figure S12), the electronic structure is significantly perturbed, and significant spin splitting was also noticed Figure S13. These results indicate that adsorption of $Li_2S_2$ changes the electronic structure in both up and down spin channels.

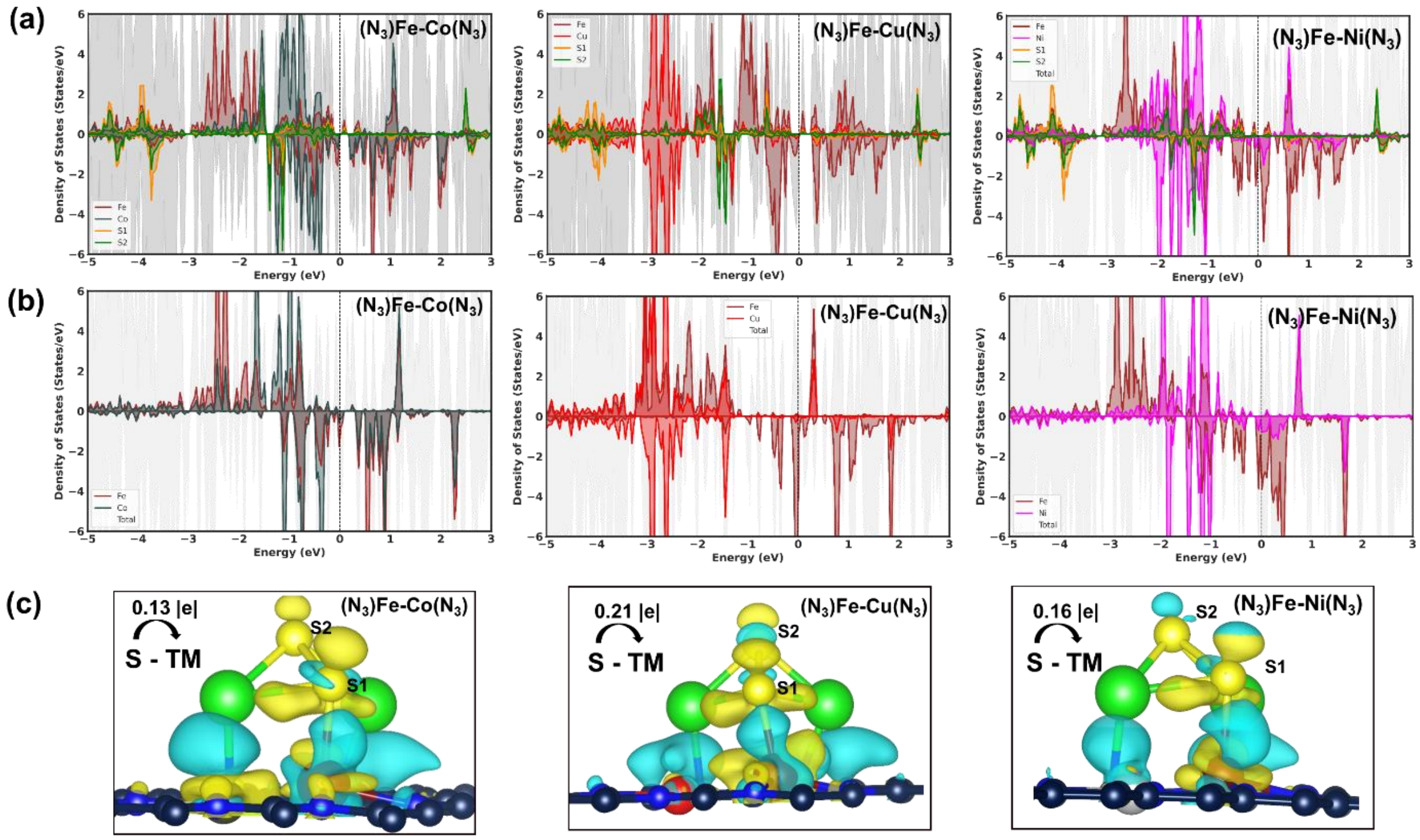

**Figure 5**: (a) Calculated PDOS (projected density of states) of $Li_2S_2$ (substrate) on the cathode hosts, (b) Calculated PDOS of both S with orbit gap and (c) Side view of charge density difference of $Li_2S_2$ adsorption on $(N_3)$ Fe-Co$(N_3)$, $(N_3)$ Fe-Cu$(N_3)$, $(N_3)$Fe-Ni$(N_3)$ respectively. The yellow and blue section represent electron accumulate and lose region. The iso surface is set to 0.003eV/Å$^3$.

To understand the relative charge transfer for sulfur atoms of $Li_2S_2$, we compared the difference between the Bader charges of adsorbed sulfur atoms. We notice that among the aforementioned DACs, higher charge transfer (relative charge on sulfur) is largest in $(N_3)$Fe-Cu$(N_3)$ (0.21 |e|), followed by $(N_3)$Fe-Ni$(N_3)$ (0.16 |e|), and lowest in $(N_3)$Fe-Co$(N_3)$ (0.13 |e|) Figure 5(c). Interestingly, the adsorption energies on $Li_2S_2$ on these DACs also follow a similar trend. Additionally, the shift in d-band centers upon $Li_2S_2$ adsorption was found to be reasonable in case of $(N_3)$Fe-Cu$(N_3)$ ($\Delta\varepsilon_d = 0.12$) however Ni and Co doped system shows nearly similar change in d-band center, $(N_3)Fe\text{-}Ni(N_3)$ ($\Delta\varepsilon_d = 0.02$) , $(N_3)Fe\text{-}Co(N_3)$ ($\Delta\varepsilon_d = 0.04$). We notice that in all DACs the sulfur atoms exhibit different bader charges, but in some DACs, such as $(N_3)Fe\text{-}Cu(N_3)$, $(N_3)$ Fe-Fe$(N_3)$, $(N_3)$ Fe-Mn$(N_3)$, $(N_3)$ Fe-Zn$(N_3)$, $(N_3)$Co-Cu$(N_3)$, $(N_3)$Co-Co$(N_3)$, $(N_3)$Co-Mn$(N_3)$, and $(N_3)$ Co-Pt$(N_3)$, these differences are more negative than 0.21|e| Table S1. On the other hand, in DACs (Fe$(N)_4$ and Co$(N)_4$), the differences in Bader charges of sulfur atoms are less negative than 0.20 |e|. These results suggest the DACs polarize the S-S bond to a greater extent than SACs Table S1.

**Catalytic Decomposition of $Li_2S$**

During the charging process, $Li_2S$ undergoes oxidative decomposition to regenerate elemental sulfur. The direct decomposition of free $Li_2S$ into LiS and a single $Li^+$ ion ($Li_2S \rightarrow LiS + Li^+ + e^-$) is associated with a high energy barrier of 3.39 eV[15], which leads to the accumulation of electrochemically inactive (dead) $Li_2S$, thereby reducing active material utilization and degrading cycle life. To investigate the kinetics of $Li_2S$ dissociation of DACs, we analyzed the bond angle[9] between Li-S-Li and the bond length of Li-S in $Li_2S$ adsorbed on DACs. In several DACs, the Li-S-Li bond angle is found to be greater than 115° Figure 6(e) while the S-S bond length was consistently elongated relative to free $Li_2S$. This elongation is attributed to the charge transfer from the M-S bond Figure 6e, suggesting that DACs may facilitate $Li_2S$ activation by lowering the decomposition barrier. To validate this assertion, we computed the Li dissociation barriers of $Li_2S$ using the CI-NEB method. Firstly, due to the high heterogeneity of the catalyst in the graphene layer, we determine the most preferred Li diffusion site by calculating the Li -atom diffusion barriers. Remarkably, several DACs, including $(N_3)$Co-Co$(N_3)$, $(N_3)$Fe-Mn$(N_3)$, $(N_3)$Fe-V$(N_3)$, $(N_3)$Fe-Ni$(N_3)$, and $(N_3)$Fe-Pt$(N_3)$, exhibited Li-S decomposition barriers below $\leq$ 1.0 eV, indicating their strong catalytic activity during charging. In contrast, DACs such as $(N_3)$Fe-Cu$(N_3)$,

$(N_3)Fe$-$Fe(N_3)$, $(N_3)Fe$-$Co(N_3)$, $(N_3)Co$–$Ni(N_3)$, $(N_3)Fe$-$Zn(N_3)$, $(N_3)Co$-$Pt(N_3)$, and $(N_3)Co$-$Cu(N_3)$ displayed higher barriers ranging from 1.07 to 1.35 eV. A systematic comparison across weak Figure 6(a), moderate Figure 6(b), and strong Figure 6(c) cathode hosts confirmed that decomposition of $Li_2S$ on DACs consistently exhibited lower activation barriers than free $Li_2S$.

To further identify optimal cathode hosts, we examined their electronic and thermodynamic descriptors, including binding energies of $Li_2S_8$ and $Li_2S$, formation energies, Gibbs free energy changes ($\Delta G$), and $Li_2S$ decomposition barriers. The comparative performance of these hosts is schematically summarized in the radar plot Figure 7(e). Among the studied systems, $(N_3)Fe$-$Ni(N_3)$ exhibits the most balanced catalytic characteristics. It possesses the lowest decomposition barrier (1.0 eV) and a highly exergonic overall reaction ($\Delta G(overall) = -1.93$ eV). The relatively small $\Delta G(max)$ value of 0.42 eV further indicates a smooth reaction pathway with minimal energetic bottlenecks. Importantly, the $\Delta G(Li_2S_4$-$Li_2S)$ value of 0.787 eV is the lowest among all DACs, suggesting enhanced kinetics for polysulfide reduction and oxidation processes. These results collectively suggest Fe-Ni as a highly efficient DAC capable of facilitating both the decomposition of $Li_2S$ during charging and the reduction of $Li_2S_4$ intermediates during discharge. Additionally, the $(N_3)Fe$-$Pt(N_3)$ system also shows promising catalytic behavior, with an exergonic $\Delta G(overall)$ of -1.96 eV and a relatively low decomposition barrier (1.08 eV). However, its higher $\Delta G(Li_2S_4$–$Li_2S)$ (1.08 eV) indicates a comparatively less favorable reduction process. DACs such as $(N_3)Fe$–$Co(N_3)$ and $(N_3)Fe$–$Cu(N_3)$ exhibit moderate catalytic performance, characterized by decomposition barriers of 1.22–1.34 eV and $\Delta G(Li_2S_4$-$Li_2S)$ values exceeding 0.85 eV, suggesting slower reaction kinetics. Conversely, $(N_3)Fe$-$Fe(N_3)$ and $(N_3)Co$-$Cu(N_3)$ display relatively high kinetic and thermodynamic barriers, implying limited catalytic efficacy.

Overall, the Fe-Ni DAC demonstrates an optimal performance due to the synergy between the two metal centers, which polarize the Li-S bond and facilitate charge redistribution within the adsorbed $Li_2S$ species. This dual-site cooperation effectively lowers the decomposition barrier and enhances polysulfide redox kinetics, positioning Fe–Ni as a superior catalyst for promoting reversible $Li_2S$ oxidation in Li-S batteries.

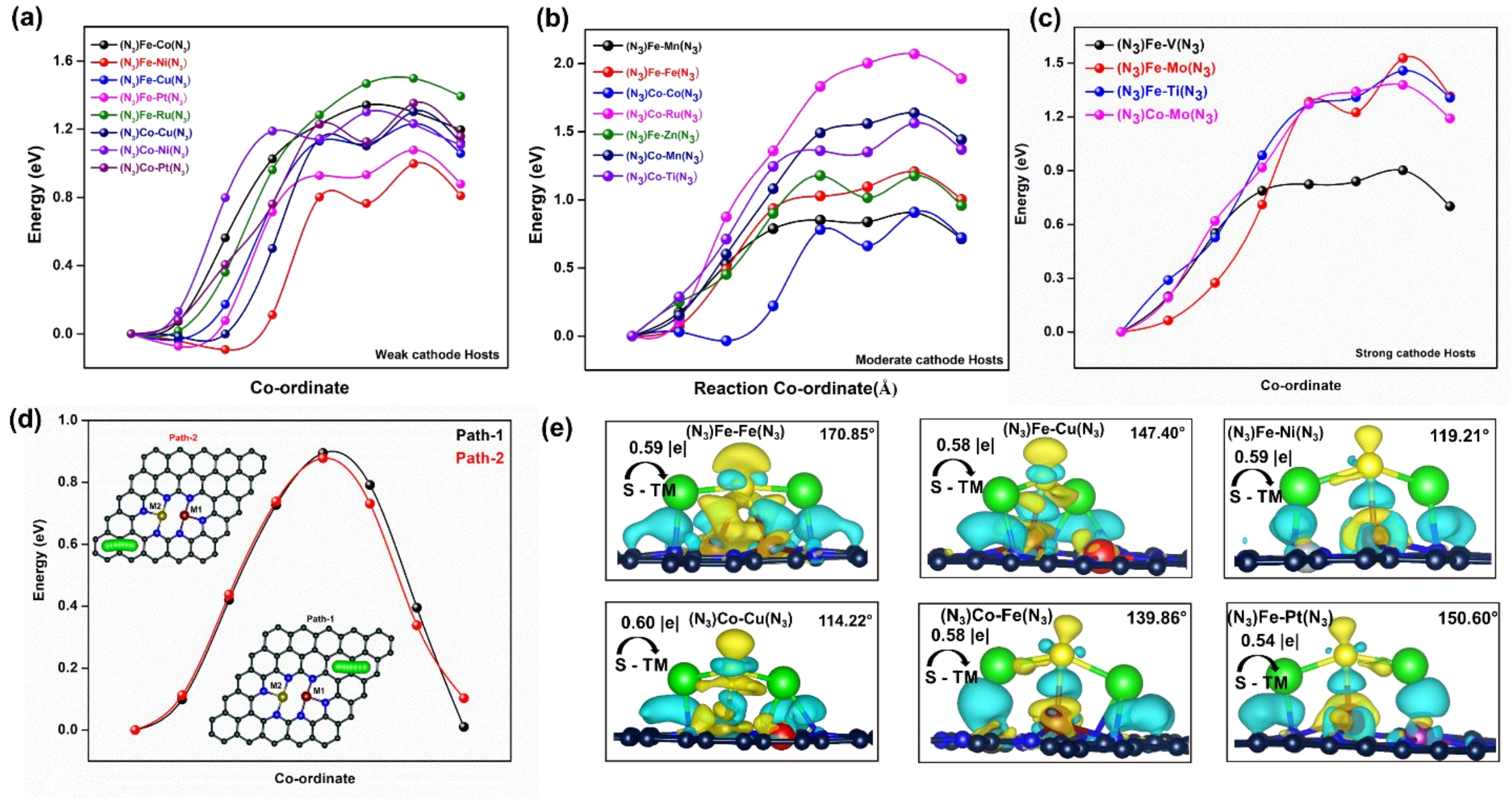

**Figure 6**: Decomposition barrier of $Li_2S$, Li-ion diffusion surface using CI-NEB and bader charges difference of $Li_2S$ on different dual metal doped graphene (a) weak cathode hosts, (b) strong cathode hosts and (c) strong cathode hosts. (d) Li-ion diffusion barrier of different paths. The inset images in panel are the detailed decomposition path of Li-ion. (e) side view of charge density difference and the bond angle of Li-S-Li of various cathode hosts mentioned in panel. The yellow and blue section represent electron accumulate and lose region. The iso surface is set to 0.003eV/$Å^3$.

## Machine Learning Investigation

Since LPS adsorption and conversion processes in Li-S batteries are primarily governed by the thermodynamic stability of intermediates, accurately predicting their Gibbs free energy ($\Delta G$) is crucial for rational catalyst design. However, computing all possible dual-metal combinations and corresponding LPS adsorbed configurations through direct DFT calculations would be computationally exhaustive. To address this challenge, we developed a machine learning (ML) workflow capable of predicting ΔG for LPSs adsorbed on $N_6$-dual-metal-doped graphene surfaces. We utilized the DFT computed dataset of 19 representative DACs and corresponding LPS ($Li_2S_n$, n = 1-8) adsorbed systems. Using this dataset, we trained multiple ML regression models to identify the correlation between free energies and easily computable structural and electronic descriptors. Prior to model training, a Pearson correlation coefficient matrix was computed to examine inter-feature relationships and identify the most relevant features. The resulting heatmap (Figure 7a) revealed that several features exhibited significant correlation with Gibbs free energy, while maintaining low mutual correlation. The identified descriptors include: (i) the average integrated crystal orbital Hamilton population (ICOHP) between both transition metals (TMs) and sulfur, (ii) the average TM-S bond length, (iii) the average Li-S bond length, (iv) the difference in electronegativity between the two dopant metals (Δχ), and (v) the number of sulfur atoms ($n_S$) in the polysulfide species. In addition to these descriptors, we also utilized a descriptor named inter-metal electron affinity[9] ($I_{EA}$) obtained by equation (4)

$$I_{Ea} = \frac{|(N_{Fe}-N_M)*(\chi_{Fe}-\chi_M)*r_{Fe}|}{\chi_S * r_M} \tag{4}$$

Where, $N_{Fe}$, $N_M$, $\boldsymbol{\chi_{Fe}}$ and $\boldsymbol{\chi_M}$ represent the numbers of d-electrons and electronegativity of Fe and doped metals, respectively. $\boldsymbol{r_{Fe}}$ and $\boldsymbol{r_M}$ are the ionic radius of Fe and doped metals, whereas $\boldsymbol{\chi_S}$ is the electronegativity of sulfur. To further refine this descriptor, we incorporated two energy-difference parameters defined as:

$$\Delta\varepsilon_{TM-S} = (\varepsilon_{TM} - \varepsilon_S)_{abs} \tag{5}$$

$$\Delta\varepsilon_{TM-N} = (\varepsilon_{TM} - \varepsilon_N)_{cat} \tag{6}$$

Subtracting these from $\boldsymbol{I_{Ea}}$ improved its correlation with the change in Gibbs free energy Figure S16.

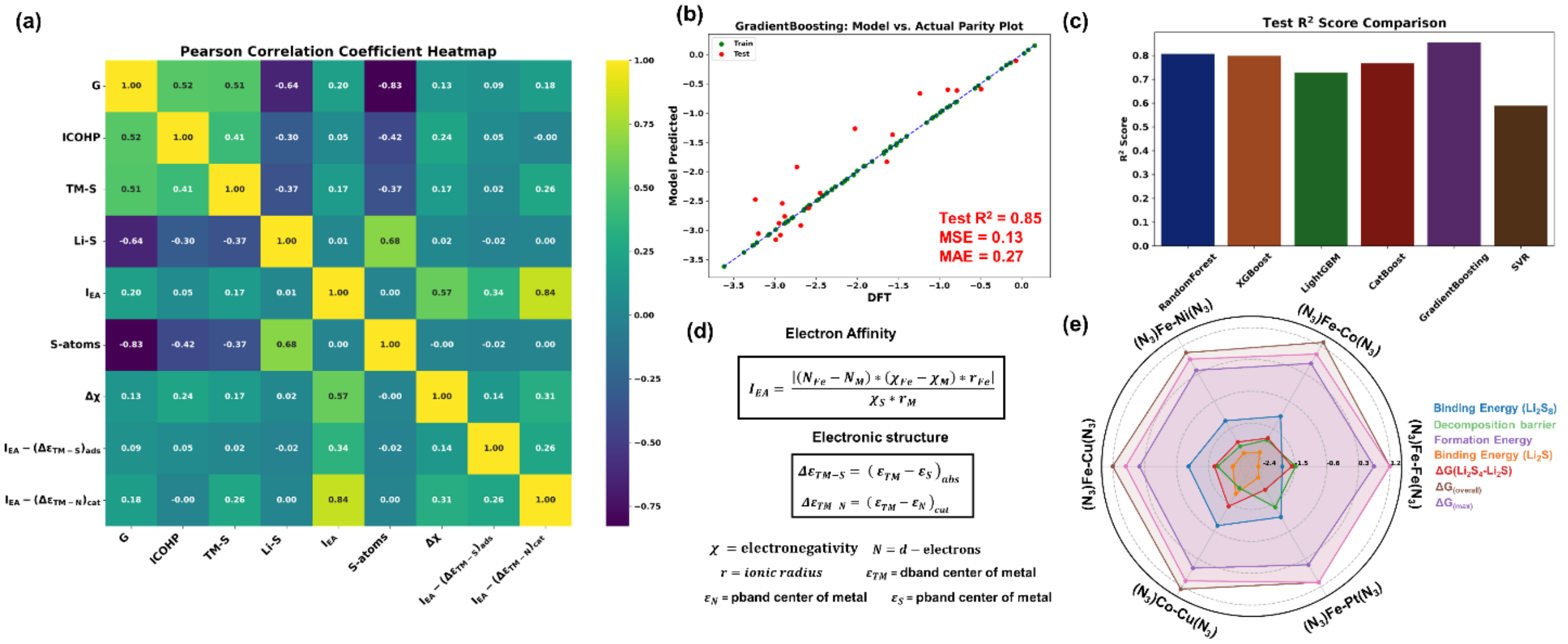

**Figure 7**: ML training and prediction for the change of Gibbs Free energy correlation with the key descriptors. (a)Heat map of the Pearson correlation coefficient matrix among the selected descriptors. (b) Comparison of the change of Gibbs Free energy with all descriptors (mentioned above) values for the training and testing data set. (c) Comparison of test $R^2$ with all the models used for ML investigation. (d) Electron affinity and Electronic structure calculations are performed using equations. (e) Radar chart of various factors for the guidance and evaluation of host materials.

To develop a robust ML model, we trained multiple regression models, including random forest regression (RFR), support vector regression (SVR), and Gradient Boosting regression (GBR), along with their advanced ensemble variants, XGBoost, LightGBM, and CatBoost. The comparative performance of these models is summarized in Figures 7(b–c). Among all, the GBR model demonstrated the best predictive capability, achieving a minimum mean squared error (MSE) of 0.12 eV, a mean absolute error (MAE) of 0.26 eV, and a test $R^2$ value of 0.85 eV. These values indicate excellent agreement between the predicted and DFT-calculated Gibbs free energies. Hyperparameter optimization was carried out using five-fold cross-validation (CV), which yielded CV($R^2$) values of 0.78 for GBR, 0.77 for XGBoost, 0.66 for LightGBM, and 0.83 for CatBoost. The RFR model also performed reasonably well (test $R^2$ = 0.80, CV($R^2$) = 0.75), while the SVR model showed comparatively weaker predictive ability (test $R^2$ = 0.58, CV($R^2$) = 0.73). The parity plots for all models Figure S17 clearly illustrate the strong correlation between predicted and reference values, confirming the reliability of our feature set and training protocol. Overall, the GBR-based model emerged as the most effective and interpretable approach for predicting Gibbs free energy variations in LPS adsorption. The identified descriptors not only capture the key physical and electronic interactions governing the Li-S conversion process but also provide valuable design principles for rational optimization of doped graphene hosts for next-generation lithium–sulfur batteries.

**Computational Details**

All spin-polarized calculations are performed within the framework of density functional theory DFT as implemented in the Vienna ab initio Simulation Package (VASP)[16]. We employed Perdew–Burke-Ernzerhof (PBE) exchange correlation[17] functional under generalized gradient approximation (GGA) as well as Grimme's DFT+D3[18] to understand the physical van der Waals (vdW) interaction between Lithium polysulfides (LiPSs) and the dual atom catalysts (DACs). The DACs graphene-based slabs shown in Figure S17 are designed with a supercell size of 5×5×1 and a vacuum of 15Å in the z direction. The formation energies ($E_f$) of all the cathode hosts are shown in Table S3. A plane wave cutoff energy of 520 eV was used to optimize all structures. The residual force for bounded and unbounded LiPSs with a slab is fully optimized with less than 0.01eV/Å by k-point sampling of 3×3×1 k-mesh using Monkhorst-Pack scheme[19]. We calculated the adsorption energy and the ratio of van der Waals interaction (vdw) of all the polysulfides using the following equations.

$$\text{Binding Energy } (E_b) = E_{(\text{host + polysulfide})} - (E_{(\text{host})} + E_{(\text{polysulfides})}) \qquad (1)$$

$$R_{vdw} = \left(\frac{E_b^{vdw} - E_b^{without\ vdw}}{E_b^{vdw}}\right) \times 100 \qquad (2)$$

Where $R_{vdw}$the ratio of van der walls interaction, $E_b^{vdw}$ , $E_b^{without\ vdw}$ is binding energy with and without van der walls interaction, respectively.

We employed the CP2K[20] code to calculate the decomposition energy barrier of $Li_2S$ using the climbing image-nudged elastic band (CI-NEB)[21] approach. Bader charges analysis is carried out to analyze the charge transfer between dual metals and LiPSs. We used the following expression to calculate the charge density difference (CDD).[22]

$$\Delta\rho = \rho_{\text{Total}} - \rho_{\text{slab}} - \rho_{\text{LiPSs}} \qquad (3)$$

Here $\rho_{\text{Total}}$ represents the charge density of slab with LiPSs and $\rho_{\text{slab}}$ represents the charge density of slab without any LiPSs and $\rho_{\text{LiPSs}}$ represents the charge density of isolated LiPSs, respectively. Furthermore, we employed an in-house built machine learning model, Precise and Accurate Configuration Evaluation (PACE), which uses machine learning interatomic potentials (MLIPs) to screen and identify low-energy configurations of adsorbates on slabs/crystals. The algorithm works by establishing a series of divisions and levels on the slab. The adsorbate is positioned on these coordinates on the slab, and single-point energy calculations are performed on different conformations of the adsorbate using the MLIP of choice. This data is used to filter out a preset number of low-energy structures, and geometry optimization using the MLIP is performed on them. The obtained optimized low-energy structure of the slab and adsorbate is further processed to density functional theory (DFT) calculations.

## Conclusion

In this work, we developed a machine-learning-accelerated DFT framework (PACE) to systematically identify lithium polysulfide ($Li_2S_n$) adsorption configuration. By utilizing Machine Learning Interatomic Potentials (MLIPs), PACE enables rapid and exhaustive identification of low-energy configurations. In addition, we investigated the detailed SRR mechanism on a series of $N_6$-coordinated dual-atom catalysts (DACs) to identify an efficient and cost-effective DACs for SRR. Our results reveal that DACs exhibit stronger and more tunable $Li_2S_n$ adsorption than SACs, due to dual-metal synergy and enhanced charge delocalization. Among the studied DACs, Fe-Ni and Fe-Pt DACs demonstrate the most favorable balance between polysulfide anchoring and $Li_2S_2$ dissociation kinetics. Furthermore, an independent machine learning regression model was developed to predict Gibbs free energies ($\Delta G$) using DFT calculation datasets with physically interpretable descriptors, including ICOHP, TM-S/Li-S bond lengths, and electronegativity difference ($\Delta\chi$). The Gradient Boosting Regression (GBR) model achieved superior predictive accuracy ($R^2$ = 0.85, MAE = 0.26 eV), providing a fast and generalizable tool for catalyst screening. Overall, this integrated PACE-ML strategy offers a powerful paradigm for the rational design of DAC-based catalysts for the development of next-generation lithium-sulfur batteries.

## Acknowledgment

M.D. and S.K. gratefully acknowledge the financial support provided by the CSIR, India, which facilitated the completion of this work. A.M.K.R. acknowledges the Department of Atomic Energy and UM-DAE-Centre for Excellence in Basic Sciences, Mumbai, for providing the DAE-DISHA fellowship. Additionally, we express our gratitude to the National Supercomputing Mission (NSM) for granting access to the computing resources of the Param Porul HPC System. This system is implemented by C-DAC and is supported by the Ministry of Electronics and Information Technology (MeitY) and the Department of Science and Technology (DST), Government of India.

## Code and Data Availability

The codes and data are available through our GitHub repositories: https://github.com/dixitmudit/ML4LiS/ and https://github.com/dixitmudit/PACE

## References:

[1] X. Ji, L. F. Nazar, *J. Mater. Chem.* **2010**, *20*, 9821.

[2] a)A. Manthiram, Y. Fu, Y.-S. Su, *Acc. Chem. Res.* **2013**, *46*, 1125-1134; b)Y.-X. Yin, S. Xin, Y.-G. Guo, L.-J. Wan, *Angew. Chem. Int. Ed.* **2013**, *52*, 13186-13200; c)A. Manthiram, S.-H. Chung, C. Zu, *Adv. Mater. or Adv Mater.* **2015**, *27*, 1980-2006; d)A. Manthiram, Y. Fu, S.-H. Chung, C. Zu, Y.-S. Su, *Chem. Rev.* **2014**, *114*, 11751-11787; e)R. Fang, S. Zhao, Z. Sun, D.-W. Wang, H.-M. Cheng, F. Li, *Adv. Mater. or Adv Mater.* **2017**, *29*, 1606823.

[3] a)T. Zhang, Z. Chen, J. Zhao, Y. Ding, *Diam. Relat. Mater.* **2018**, *90*, 72-78; b)P. G. Bruce, S. A. Freunberger, L. J. Hardwick, J.-M. Tarascon, *Nat. Mater.* **2012**, *11*, 19-29; c)H.-J. Peng, J.-Q. Huang, X.-B. Cheng, Q. Zhang, *Adv. Energy Mater* **2017**, *7*, 1700260; d)S.-H. Chung, A. Manthiram, in *Encyclopedia of Electrochemistry*, 2020, pp. 1-36; e)Z. W. Seh, Y. Sun, Q. Zhang, Y. Cui, *Chem. Soc. Rev.* **2016**, *45*, 5605-5634; f)A. Eftekhari, D.-W. Kim, *J. Mater. Chem.* **2017**, *5*, 17734-17776.

[4] a)X. Tao, F. Chen, Y. Xia, H. Huang, Y. Gan, X. Chen, W. Zhang, *ChemComm* **2013**, *49*, 4513-4515; b)H. Yuan, T. Liu, Y. Liu, J. Nai, Y. Wang, W. Zhang, X. Tao, *Chem Sci.* **2019**, *10*, 7484-7495; c)D. Guo, X. Zhang, M. Liu, Z. Yu, X. a. Chen, B. Yang, Z. Zhou, S. Wang, *Adv. Funct. Mater.* **2022**, *32*, 2204458.

[5] a)B. Qiao, A. Wang, X. Yang, L. F. Allard, Z. Jiang, Y. Cui, J. Liu, J. Li, T. Zhang, *Nat. Chem.* **2011**, *3*, 634-641; b)Y. Zhang, J. Zhao, H. Wang, B. Xiao, W. Zhang, X. Zhao, T. Lv, M. Thangamuthu, J. Zhang, Y. Guo, J. Ma, L. Lin, J. Tang, R. Huang, Q. Liu, *Nat. Comm.* **2022**, *13*, 58; c)K. Ye, M. Hu, Q.-K. Li, Y. Han, Y. Luo, J. Jiang, G. Zhang, *J. Phys. Chem. Lett.* **2020**, *11*, 3962-3968; d)S. Mukherjee, X. Yang, W. Shan, W. Samarakoon, S. Karakalos, D. A. Cullen, K. More, M. Wang, Z. Feng, G. Wang, G. Wu, *Small Methods* **2020**, *4*, 1900821; e)X. Sun, Y. Qiu, B. Jiang, Z. Chen, C. Zhao, H. Zhou, L. Yang, L. Fan, Y. Zhang, N. Zhang, *Nat. Commun.* **2023**, *14*, 291.

[6] a)L. Zhang, D. Liu, Z. Muhammad, F. Wan, W. Xie, Y. Wang, L. Song, Z. Niu, J. Chen, *Adv. Mater. or Adv Mater.* **2019**, *31*, 1903955; b)B.-Q. Li, L. Kong, C.-X. Zhao, Q. Jin, X. Chen, H.-J. Peng, J.-L. Qin, J.-X. Chen, H. Yuan, Q. Zhang, J.-Q. Huang, *InfoMat* **2019**, *1*, 533-541; c)G. Zhou, S. Zhao, T. Wang, S.-Z. Yang, B. Johannessen, H. Chen, C. Liu, Y. Ye, Y. Wu, Y. Peng, C. Liu, S. P. Jiang, Q. Zhang, Y. Cui, *Nano Lett.* **2020**, *20*, 1252-1261.

[7] K. Zhang, Z. Chen, R. Ning, S. Xi, W. Tang, Y. Du, C. Liu, Z. Ren, X. Chi, M. Bai, C. Shen, X. Li, X. Wang, X. Zhao, K. Leng, S. J. Pennycook, H. Li, H. Xu, K. P. Loh, K. Xie, *ACS Appl. Mater. Interfaces.* **2019**, *11*, 25147-25154.

[8] Z. Han, S. Zhao, J. Xiao, X. Zhong, J. Sheng, W. Lv, Q. Zhang, G. Zhou, H.-M. Cheng, *Adv. Mater. or Adv Mater.* **2021**, *33*, 2105947.

[9] Y. Wang, C. Xu, B. Li, M. Tian, M. Liu, D. Zhu, S. Dou, Q. Zhang, J. Sun, *ACS Nano* **2024**, *18*, 34858-34869.

[10] a)X. Zhang, T. Yang, Y. Zhang, X. Wang, J. Wang, Y. Li, A. Yu, X. Wang, Z. Chen, *Adv. Mater. or Adv Mater.* **2023**, *35*, 2208470; b)Z. Zhu, Y. Zeng, Z. Pei, D. Luan, X. Wang, X. W. Lou, *Angew. Chem. Int. Ed.* **2023**, *62*, e202305828; c)J. Wu, Y. Feng, Y. Chen, T. Fan, Y. Li, *J. Mater. Chem.* **2023**, *11*, 12025-12033; d)T. Zhang, D. Luo, H. Xiao, X. Liang, F. Zhang, H. Zhuang, M. Xu, W. Dai, S. Qi, L. Zheng, Q. Gao, *Small* **2024**, *20*, 2306806.

[11] a)V. L. Deringer, M. A. Caro, G. Csányi, *Adv. Mater. or Adv Mater.* **2019**, *31*, 1902765; b)M. Kulichenko, B. Nebgen, N. Lubbers, J. S. Smith, K. Barros, A. E. A. Allen, A. Habib, E. Shinkle, N. Fedik, Y. W. Li, R. A. Messerly, S. Tretiak, *Chem. Rev.* **2024**, *124*, 13681-13714.

[12] A. Pacini, M. Ferrario, M. C. Righi, *J. Chem. Theory Comput.* **2025**, *21*, 7102-7110.

[13] S. Feng, Z.-H. Fu, X. Chen, Q. Zhang, *InfoMat* **2022**, *4*, e12304.

[14] a)G. Yilmaz, T. Yang, Y. Du, X. Yu, Y. P. Feng, L. Shen, G. W. Ho, *Adv. Sci.* **2019**, *6*, 1900140; b)R. S. Assary, L. A. Curtiss, J. S. Moore, *J. Phys. Chem. C* **2014**, *118*, 11545-11558; c)J. Shen, Z. Wang, X. Xu, Z. Liu, D. Zhang, F. Li, Y. Li, L. Zeng, J. Liu, *Adv. Energy Sustain. Res.,* **2021**, *2*.

[15] D. Wang, F. Li, R. Lian, J. Xu, D. Kan, Y. Liu, G. Chen, Y. Gogotsi, Y. Wei, *ACS Nano* **2019**, *13*, 11078-11086.

[16] a)G. Kresse, J. Furthmüller, *PhysRevB.* **1996**, *54*, 11169-11186; b)G. Kresse, J. Hafner, *PhysRevB.* **1994**, *49*, 14251-14269; c)G. Kresse, J. Hafner, *PhysRevB.* **1993**, *47*, 558-561; d)G. Kresse, D. Joubert, *PhysRevB.* **1999**, *59*, 1758-1775.

[17] J. P. Perdew, K. Burke, M. Ernzerhof, *PhysRevLett.* **1996**, *77*, 3865-3868.

[18] a)A. D. Becke, E. R. Johnson, *J. Chem. Phys.* **2005**, *123*, 154101; b)H. Schröder, A. Creon, T. Schwabe, *J. Chem. Theory Comput.* **2015**, *11*, 3163-3170; c)S. Grimme, S. Ehrlich, L. Goerigk, *J. Comput. Chem.* **2011**, *32*, 1456-1465.

[19] H. J. Monkhorst, J. D. Pack, *PhysRevB.* **1976**, *13*, 5188-5192.

[20] J. VandeVondele, M. Krack, F. Mohamed, M. Parrinello, T. Chassaing, J. Hutter, *Comput. Phys. Commun.* **2005**, *167*, 103-128.

[21] G. Henkelman, B. P. Uberuaga, H. Jónsson, *J. Chem. Phys.* **2000**, *113*, 9901-9904.

[22] G. Henkelman, A. Arnaldsson, H. Jónsson, *Comput. Mater. Sci.* **2006**, *36*, 354-360.

**TOC graphic**

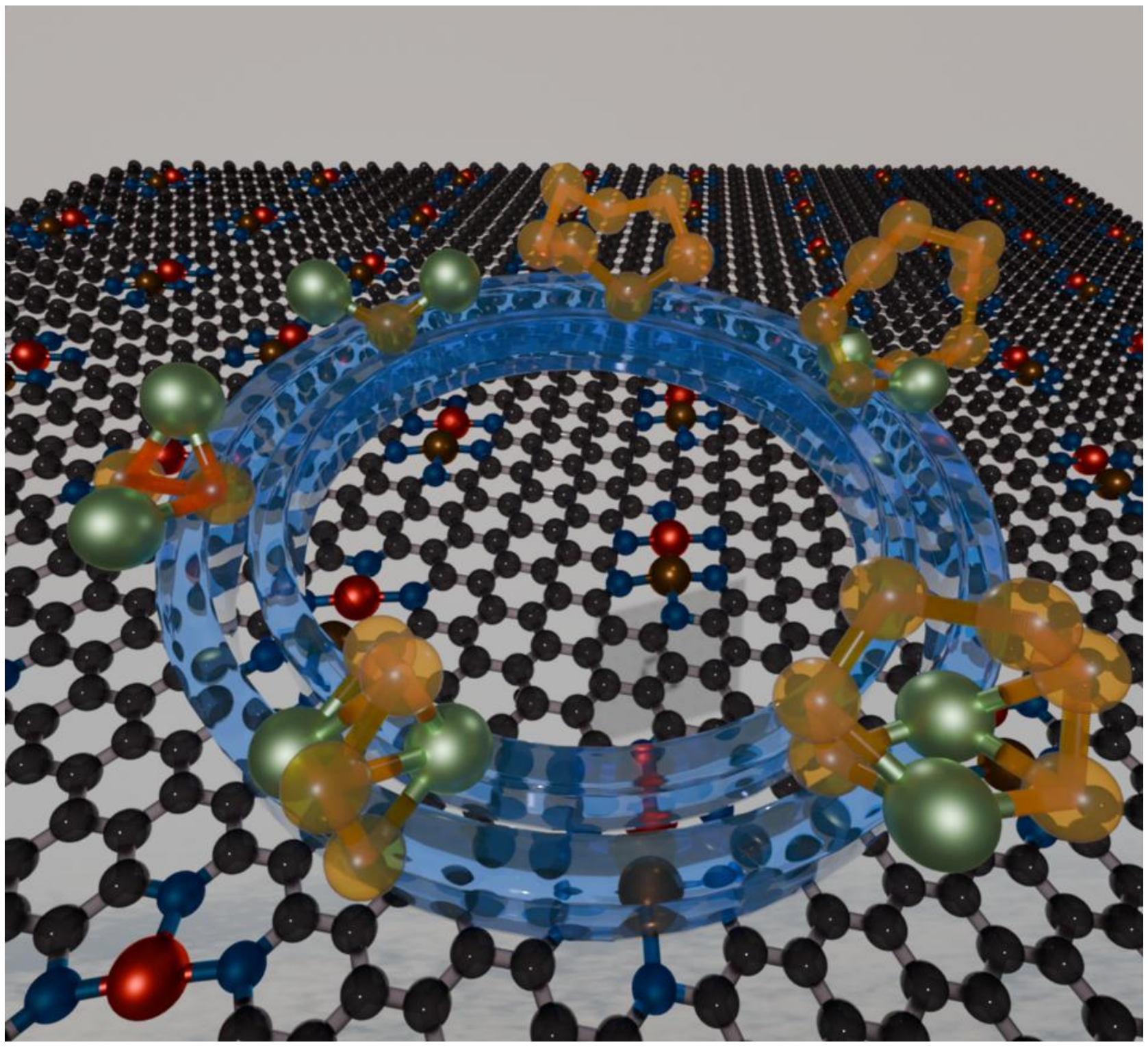

By utilizing Machine Learning Interatomic Potentials (MLIPs), PACE enables rapid and exhaustive identification of low-energy configurations. The SRR mechanisms on various $N_6$-coordinated DACs were systematically explored to screen out the most efficient and economical catalysts.